\documentclass[floatfix,aps,twocolumn,preprintnumbers,amsmath,amssymb,nofootinbib,superscriptaddress,prl]{revtex4}

\usepackage{graphicx,color}
\usepackage{amsmath,amssymb,bm}
\usepackage{dcolumn}

\newcommand{\np}{{\bf p}}
\newcommand{\nq}{{\bf q}}

\def\XXint#1#2#3{{\setbox0=\hbox{$#1{#2#3}{\int}$}
     \vcenter{\hbox{$#2#3$}}\kern-.5\wd0}}

\def\1{\'{\i}}

\begin{document}

\title{Superscaling analysis of quasielastic electron scattering 
with relativistic effective mass }

\author{J.E. Amaro}\email{amaro@ugr.es} \affiliation{Departamento de
  F\'{\i}sica At\'omica, Molecular y Nuclear \\ and Instituto Carlos I
  de F{\'\i}sica Te\'orica y Computacional \\ Universidad de Granada,
  E-18071 Granada, Spain.}
  
\author{E. Ruiz
  Arriola}\email{earriola@ugr.es} \affiliation{Departamento de
  F\'{\i}sica At\'omica, Molecular y Nuclear \\ and Instituto Carlos I
  de F{\'\i}sica Te\'orica y Computacional \\ Universidad de Granada,
  E-18071 Granada, Spain.} 

\author{I. Ruiz Simo}\email{ruizsig@ugr.es} \affiliation{Departamento de
  F\'{\i}sica At\'omica, Molecular y Nuclear \\ and Instituto Carlos I
  de F{\'\i}sica Te\'orica y Computacional \\ Universidad de Granada,
  E-18071 Granada, Spain.}

\date{\today}

\begin{abstract}
\rule{0ex}{3ex} 

We provide a parametrization of a new phenomenological scaling
function obtained from a chi-square fit to a selected set of (e,e')
cross section data expanding a band centered around the quasielastic
peak.  We start from a re-analysis of quasielastic electron scattering
from nuclear matter within the relativistic mean field model.  The
cross section depends on the relativistic effective mass of the
nucleon, $m_N^*$, and it scales with respect to a new scaling
variable, $\psi^*$.  This suggests a new superscaling approach with
effective mass (SuSAM*) for predicting quasielastic cross sections
within an uncertainty band. The model reproduces previously established
results on the enhancement of the transverse response function as
compared to the traditional relativistic Fermi gas.

\end{abstract}



\maketitle

\section{Introduction}

The description of quasielastic electron scattering cross section is
still an open problem in theoretical nuclear physics. The recent
neutrino experiments with accelerators have emphasized the importance
of a global precise description of lepton scattering from nuclei at
intermediate energies \cite{Nomad09,Agu10,Agu13,Fio13,Abe13}.
Although a great deal of progress has been achieved with models based
on first principles description of the nuclear system \cite{Lov16},
alternative approaches based on the spectral function
\cite{Ank15,Roc16} or the shell model \cite{Pan15}, to mention some
recent studies, have been put forward. As a general rule, the models
cannot provide yet a complete description of the whole set of $(e,e')$
data at the full range of kinematics needed, specially for high
momentum and energy transfers, where a relativistic description becomes
mandatory \cite{Ama02,Ama05}.  Not to mention that there are still
basic issues like gauge invariance that are not easy to control and
generate well known ambiguities.

In addition to the relativistic corrections in the kinematics and in
the current operator, the importance of relativistic corrections
stemming from the dynamics has been emphasized from a fully
relativistic mean field calculation \cite{Cab07}. In those studies the
combination of scalar and vector relativistic potentials naturally
produce an enhancement of the lower components of relativistic nucleon
wave functions \cite{Udi99} and correspondingly an enhancement of the
transverse response function.  This is a genuine relativistic
dynamical effect, and it goes away, for example, after a
semirelativistic approximation where the lower components are
neglected or projected out \cite{Ama05}.  It is known that the
transverse cross section is larger than the predictions of the
independent nucleon model. While the transverse enhancement has
primarily been attributed to multi-nucleon processes via
meson-exchange currents and $\Delta$ excitation \cite{Bod14}, it
should be noted that this enhancement can also be regarded partly as
due to relativity.

A major difficulty in the description of the inclusive $(e,e')$ cross
section is that it is the result of contributions from many unseen
processes and interferences that cannot be disentangled easily. From 
simplistic viewpoint the cross section can be regarded as the sum of
nucleon knockout plus multinucleon knockout plus pion-nucleon emission
plus additional inelastic processes in the deep region.  Both 1p-1h
and 1$\pi$1p-1h are contaminated by 2p-2h and cannot be unambiguously
separated from it \cite{Gil97,Sim16,Nie17}.  From the theoretical
point of view the one-nucleon knockout process generates the
quasielastic peak. But experimentally this peak can only be isolated
from the data of the longitudinal response function for moderate
momentum transfer where meson-exchange currents and pion emission are
predominantly transverse \cite{Don99a,Don99b}.

Scaling ideas have prompted studies which are promising phenomenological
alternatives complementary to the theoretical microscopic models
\cite{Alb88}.  The superscaling approach (SuSA)
\cite{Don99a,Don99b,Mai02,Ama04} exploited the scaling properties of
the reduced longitudinal response function (divided by the
corresponding single-nucleon structure function) when plotted against
an appropriate scaling variable $\psi'$. This allowed to implement a
phenomenological longitudinal scaling function $f_{L}(\psi')$ fitted
to electron data. This scaling function embodies implicitly all the
genuinely quasielastic nuclear physics processes.  Thus any model
aiming to describe the quasielastic reaction should be able to
describe $f_L$.

All the processes violating scaling contribute mainly to the
transverse response.  Within the SuSA approach the ``quasielastic''
part of the transverse response function was computed by assuming the
same scaling properties as the longitudinal response, and that a
transverse (unknown) scaling function $f_T$ exists. In the original
SuSA approach it was simply assumed that $f_T=f_L$ \cite{Ama04} and
this allowed to construct a manageable model to predict neutrino cross
sections from the $(e,e')$ data. Although such an assumption was not
based on data, semirelativistic models like that of ref. \cite{Ama05}
give in fact $f_T \simeq f_L$.

The relativistic mean field (RMF) framework to finite nuclei
reproduces the experimental $f_L$ function rather well and therefore
this model can be used to test the validity of the $f_T=f_L$
assumption under relativistic dynamics.  This model is based on the
Dirac-Hartree approximation of ref. \cite{Hor91}.  In \cite{Cab07} it
was actually found that in the RMF model $f_T > f_L$; the
precise value of $f_T$ depends on the treatment of the off-shell
ambiguities of the current operator.  Under the CC2 prescription of de
Forest \cite{For83}, $f_T$ is about 20\% larger than $f_L$, while it
is almost twice as large for CC1 \cite{Cab07}.  The CC2 results for
the $(e,e')$ cross section seem more reasonable, and so this was the
prescription used in the recent upgrade SuSA-v2 \cite{Gon14}. This
new model includes nuclear effects which are theoretically-inspired by
the RMF by using a transverse scaling function $f_T$ which is
different from $f_L$. and that also has an additional dependence on
the momentum transfer $q$. Therefore the SuSA-v2 results do not scale
anymore, although the model conserves the 'scaling' name by tradition.

In this work we explore a new scaling approach to describe, in a wide
kinematical range, the $(e,e')$ data by minimizing undesirable
contaminations from inelastic scattering or other effects beyond the
quasielastic conditions. We proceed by exploiting the scaling
formalism and, at the same time, the proved good properties of the
relativistic mean field, which already includes by construction the
transverse response enhancement.  Our goal is to return to the
description of the quasielastic peak with only one phenomenological
scaling function, $f^*(\psi^*)$, by fitting a subset of conveniently
selected data.  Thus, we expect that the double differential cross
section improves over the SuSA model \cite{Don99a,Gon14}. We remind
that the description of the SuSA model is not fully satisfactory
because only the longitudinal response function was fitted and not the
cross section. The SuSAv2 was an improvement by using a theoretical
transverse scaling function coming from the RMF but at the cost of
violating scaling \cite{Gon14}. In our approach, however, we maintain
the scaling with respect to a new scaling variable $\psi^*$ to be
defined shortly.

In a previous work \cite{Ama15} we started exploring the $\psi^*$
scaling idea in the context of the RMF for nuclear matter.  In that
study we obtained the best value of the effective mass
\begin{equation}
M^* = \frac{m_N^*}{m_N} = 0.8.
\end{equation}
This value provides the best scaling behavior of the data with a
large fraction of data concentrated around the universal scaling
function of the relativistic Fermi gas
\begin{equation}  
f_{\rm RFG}(\psi^*) = \frac{3}{4}(1-\psi^*{}^2) \theta(1-\psi^*{}^2)
\label{f}
\end{equation}
The $\psi^*$ variable was inspired by the mean field theory, that
provides a reasonable description of the quasielastic response
function \cite{Ros80,Ser86}. In the interacting RFG the vector and
scalar potentials generate an effective mass $m^*_N$ for the nucleon
in the medium.  Our present approach, called SuSAM* (super scaling
approach with $M^*$), enjoys the good features of the RMF in nuclear
matter.  It keeps gauge invariance (that SuSA violates because it
introduces an energy shift to account for separation energy) and
describes the dynamical enhancement of both the lower components of
the relativistic spinors and the transverse response function.

In ref. \cite{Ama15} the effective mass was randomly modified around
the mean value $0.8\pm 0.1$, approximately simulating the dispersion
band of the real data. In the present work we instead fit a selection
of experimental data which are considered ``true'' quasielastic based
on a data density criterium. Our main goal is to provide a simple fit
of the new phenomenological $\psi^*$-scaling function, $f^*(\psi^*)$
as the sum of two Gaussian functions, and similar fits for the
dispersion band as well. This simple formula allows to predict the
quasielastic cross section for arbitrary kinematics together with a
uncertainty band, providing the maximum information from the scaling
properties of the available $(e,e')$ data, with few parameters. The
uncertainty bands describe about 1000 ``quasielastic'' data out of the
$\sim 2500$ existing data for $^{12}$C. The data that lie outside the
uncertainty band, are those generated by inelastic processes or low
energy nuclear effects that break scaling explicitly.

\section{Formalism}

We follow  the notation introduced in Ref. \cite{Ama15}.
We assume that an incident electron scatters off a nucleus 
transferring momentum $\nq$ and energy $\omega$

The quasielastic cross section is 
\begin{equation}
\frac{d\sigma}{d\Omega'd\epsilon'}
= \sigma_{\rm Mott}
\left\{  
v_L R_L + 
v_T  R_T  \right\}
\end{equation}
where $\sigma_{\rm Mott}$ is the Mott cross section, 
and $\theta$ the scattering
angle.  $R_L(q,\omega)$ and $R_T(q,\omega)$ are 
the nuclear longitudinal and transverse response functions, respectively. 
The four-momentum transfer is $Q^2=\omega^2-q^2 <0$.
Finally  the kinematical factors $v_L,v_T$ are defined by 
\begin{eqnarray}
v_L &=& 
\frac{Q^4}{q^4} \\
v_T &=&  
\tan^2\frac{\theta}{2}-\frac{Q^2}{2q^2}.
\end{eqnarray}

\subsection{RMF in nuclear matter}

The SuSAM* model is inspired by the RMF in nuclear matter
which we summarize here.
In this model we consider single-nucleon excitations 
with initial nucleon energy $E=\sqrt{\np^2+m_N^*{}^2}$ in the mean
field. The final momentum of the nucleon is $\np'=\np+\nq$ and its
energy is $E'=\sqrt{\np'{}^2+m_N^*{}^2}$. Note that initial and final
nucleons have the same effective mass $m_N^*$.  
The Fermi momentum $k_F=225$ MeV/c for $^{12}$C.  
The nuclear response functions can be written in 
the  factorized form for $K=L,T$
\begin{equation}
R_K  =   r_K f^*(\psi^*), \label{factorization}  
\end{equation}
where $r_L$ and $r_T$ are the single-nucleon contribution, 
taking into account the Fermi motion
\begin{equation}
r_K = \frac{\xi_F}{m^*_N \eta_F^3 \kappa} (Z U^p_K+NU^n_K)
\label{single} 
\end{equation}
and $f^*(\psi^*)$ is the scaling function, given by Eq. (2).
It depends only on the new scaling variable $\psi^*$, that 
is the minimum kinetic 
energy of the initial nucleon divided by the kinetic Fermi energy.
The minimum energy
allowed for a nucleon inside the nucleus to absorb the virtual photon
(in units of $m_N^*$) is
\begin{equation}
\epsilon_0={\rm Max}
\left\{ 
       \kappa\sqrt{1+\frac{1}{\tau}}-\lambda, \epsilon_F-2\lambda
\right\},
\end{equation}
where we have introduced the dimensionless variables 
\begin{eqnarray}
\lambda  &=& \omega/2m_N^*, \\
\kappa  & = & q/2m_N^*,\\
\tau & = & \kappa^2-\lambda^2, \\
\eta_F & = &  k_F/m_N^*,\\
\xi_F & = & \sqrt{1+\eta_F^2}-1,\\
\epsilon_F &=& \sqrt{1+\eta_F^2}, 
\end{eqnarray}
Usually \cite{Alb88} these
variables are defined dividing by the nucleon mass $m_N$ instead
of $m_N^*$.
The definition of the scaling variable is
\begin{equation}
\psi^* = \sqrt{\frac{\epsilon_0-1}{\epsilon_F-1}} {\rm sgn} (\lambda-\tau)
\end{equation}
$\psi^*$ is negative to the left of the quasielastic peak ($\lambda < \tau$)
and positive on the right side.

The single nucleon response functions are
\begin{eqnarray}
U_L &=& \frac{\kappa^2}{\tau}
\left[ (G^*_E)^2 + \frac{(G_E^*)^2 + \tau (G_M^*)^2}{1+\tau}\Delta \right]
\\
U_T &=& 2\tau  (G_M^*)^2 + \frac{(G_E^*)^2 + \tau (G_M^*) ^2}{1+\tau}\Delta
\end{eqnarray}
where the quantity $\Delta$ has been introduced
\begin{equation}
\Delta= \frac{\tau}{\kappa^2}\xi_F(1-\psi^*{}^2)
\left[ \kappa\sqrt{1+\frac{1}{\tau}}+\frac{\xi_F}{3}(1-\psi^*{}^2)\right].
\end{equation}
One of the consequences  of the present RMF approach 
is that the electric and magnetic form factors, are modified in the medium due
to the effective mass according to
\begin{eqnarray}
G_E^*  &=&  F_1-\tau \frac{m^*_N}{m_N} F_2 \\
G_M^*  &=& F_1+\frac{m_N^*}{m_N} F_2.  \label{GM}
\end{eqnarray}
We use here the successful CC2 prescription of the electromagnetic
nucleon current, that reproduces the experimental superscaling
function \cite{Cab07}.  Using the CC1 operator obtained through the
Gordon reduction produces the same effects as in the RMF of
ref. \cite{Cab07}.  The same modification of form factors in the
medium was explored in ref. \cite{Bar98}, but the $\psi^*$-scaling was
not investigated in that context.  For the free Dirac and Pauli form
factors, $F_1$ and $F_2$, we use the Galster parametrization.

Note that our formalism generalizes the conventional SuSA formulae. In
fact for $M^*=1$ we recover the SuSA results and definitions.

The factorization of Eq. (\ref{factorization}) implies that the cross
section in the RMF for nuclear matter 
also factorizes as the product of the scaling
function times a single nucleon cross section. In this approach the
scaling function is a parabola.

\subsection{The SuSAM*}

In the SuSAM* approach we compute the
cross section by using the above formulas, but replacing the scaling
function (2) by a phenomenological one 
which we extract from the experimental
data.

\begin{figure}
\includegraphics[width= 8cm, bb=160 360 440 780]{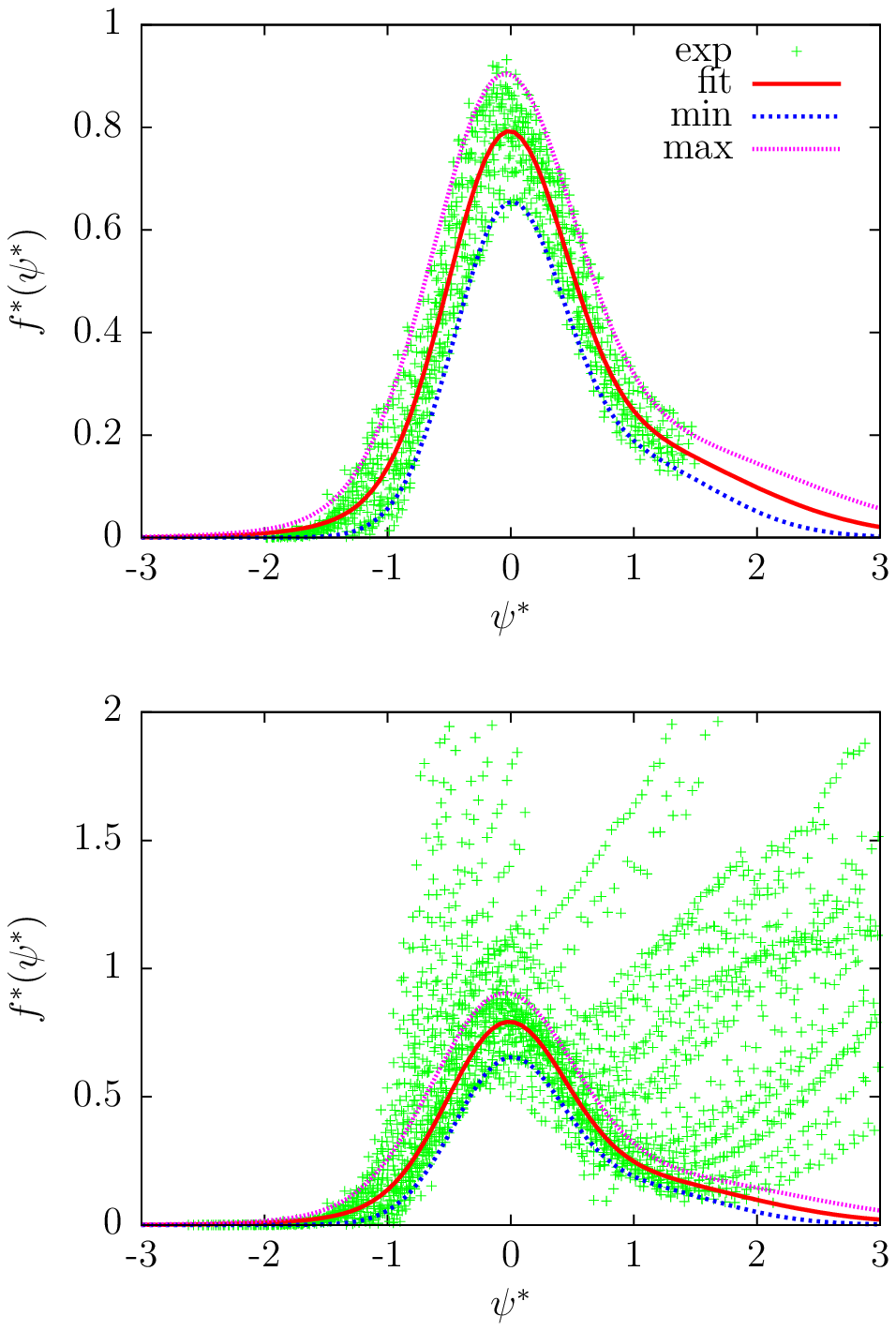}
\caption{Top panel: Phenomenological $M^*$-scaling function
  $f^*(\psi^*)$ and its uncertainty band, $f^*_{min}< f^* < f^*_{max}$, for
  $^{12}$C, obtained from a fit to the experimental data. Only data
  with density $n\geq 25$ inside a circle with radius $r=0.1$ have
  been included in the fit.  Bottom panel: comparison of the present
  fit to the bulk set of world data. Data are from ref. 
\cite{Ben08,archive,archive2} }
\label{fit}
\end{figure}

\begin{figure}
\includegraphics[width= 8cm, bb=160 120 400 780]{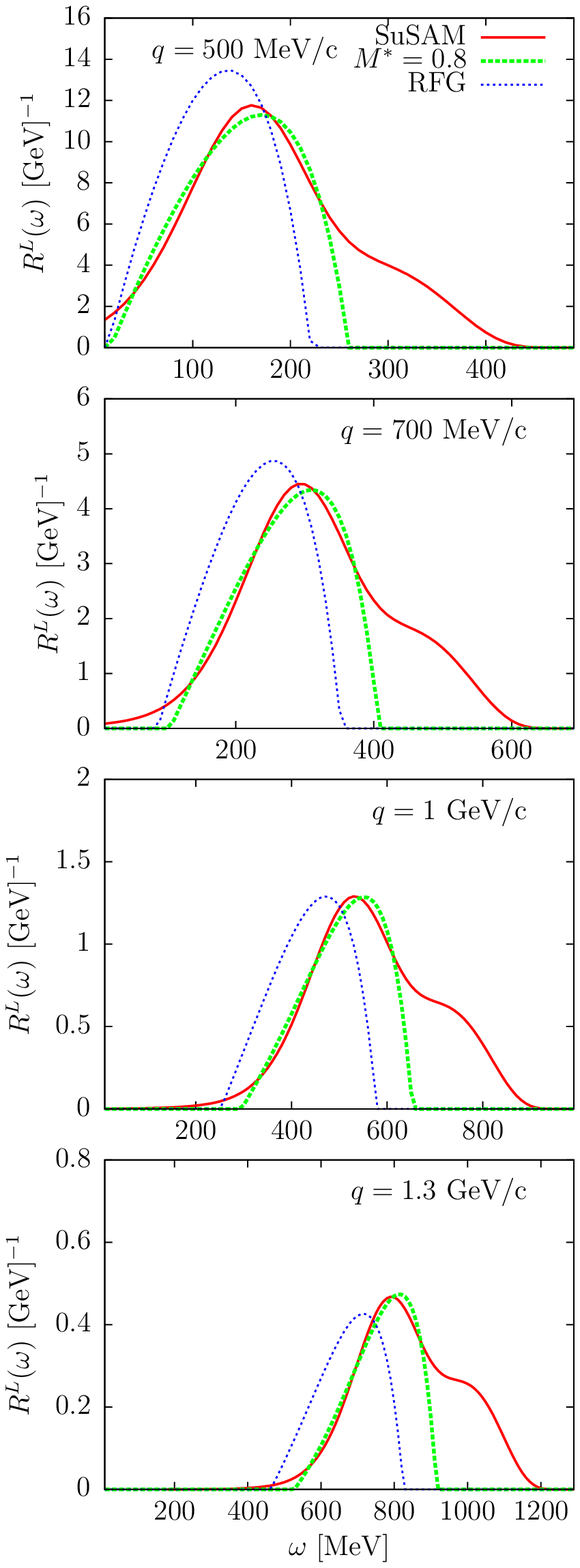}
\caption{ Longitudinal response function of $^{12}$C in the SuSAM
  model, for several values of the momentum transfer.  The
  relativistic Fermi gas results for effective mass $M^*=1$ and 0.8 are
  also shown. The Fermi momentum is $k_F=225$ MeV/c.  }
\label{RL}
\end{figure}

\begin{figure}
\includegraphics[width= 8cm, bb=160 120 400 780]{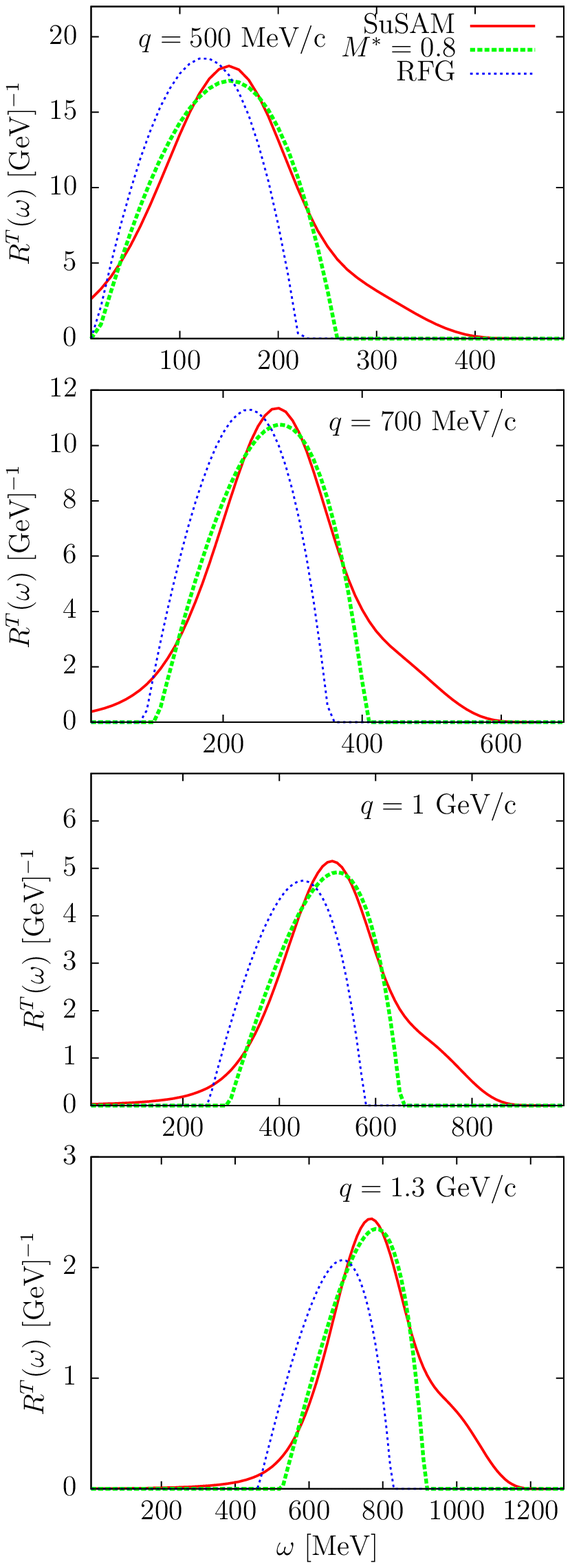}
\caption{ Transverse response function of $^{12}$C in the SuSAM
  model, for several values of the momentum transfer.  The
  relativistic Fermi gas results for effective mass $M^*=1$ and 0.8 are
  also shown. The Fermi momentum is $k_F=225$ MeV/c.  }
\label{RT}
\end{figure}

\begin{figure}
\includegraphics[width= 8cm, bb=160 120 400 780]{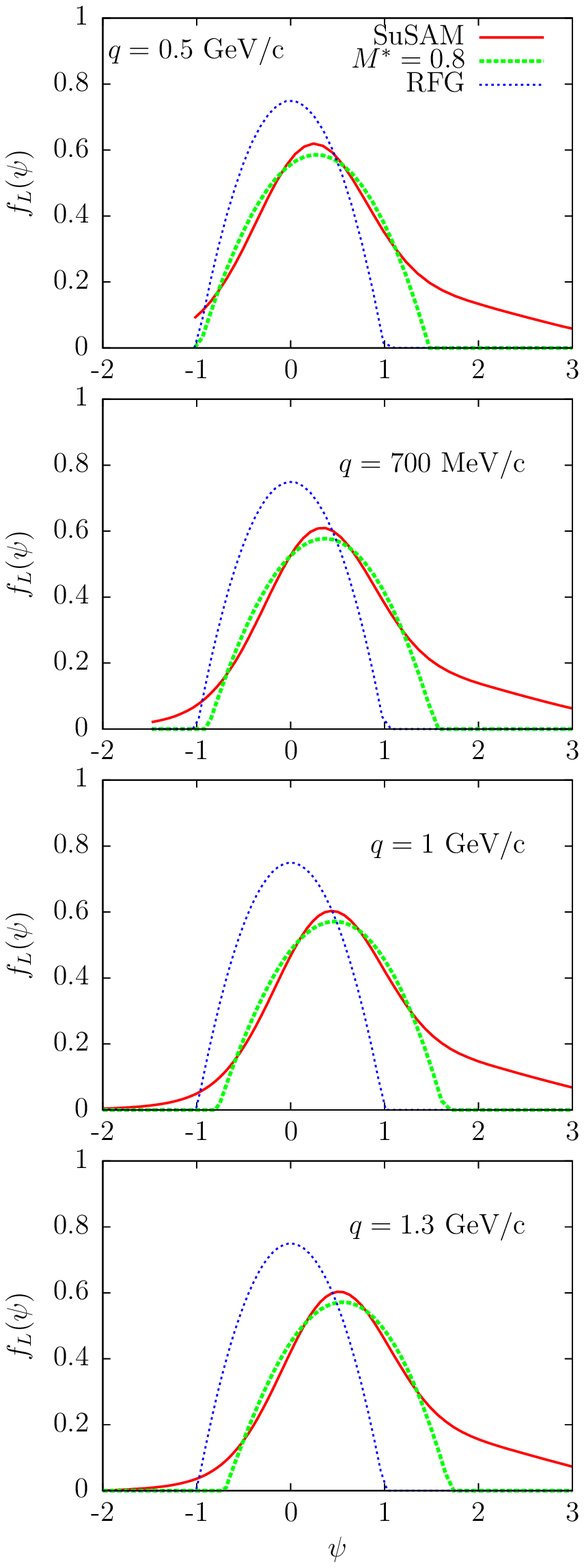}
\caption{ Traditional longitudinal scaling function $f_L(\psi)$ of $^{12}$C 
in the SuSAM
  model, for several values of the momentum transfer.  The
  relativistic Fermi gas results for effective mass $M^*=1$ and 0.8 are
  also shown. The Fermi momentum is $k_F=225$ MeV/c.  }
\label{fL}
\end{figure}

\begin{figure}
\includegraphics[width= 8cm, bb=160 120 400 780]{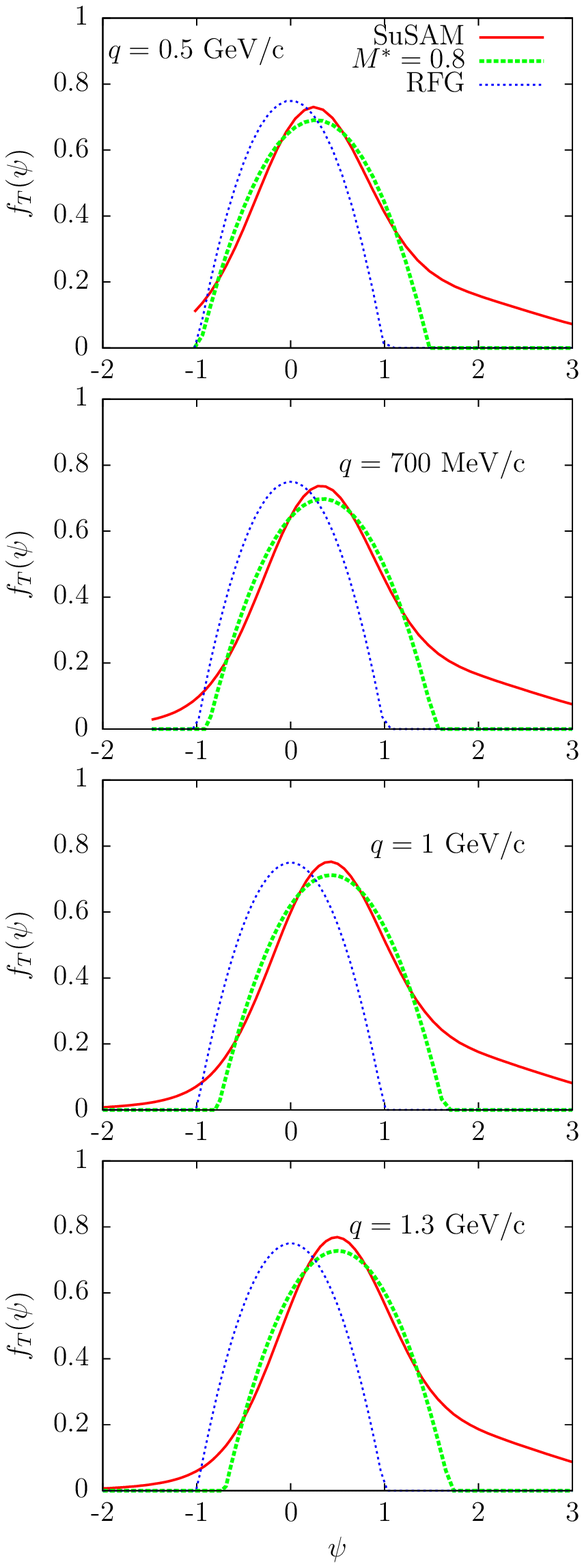}
\caption{ Traditional transverse scaling function $f_T(\psi)$ of $^{12}$C 
in the SuSAM
  model, for several values of the momentum transfer.  The
  relativistic Fermi gas results for effective mass $M^*=1$ and 0.8 are
  also shown. The Fermi momentum is $k_F=225$ MeV/c.  }
\label{fT}
\end{figure}

\begin{figure}
\includegraphics[width= 8cm, bb=130 340 430 780]{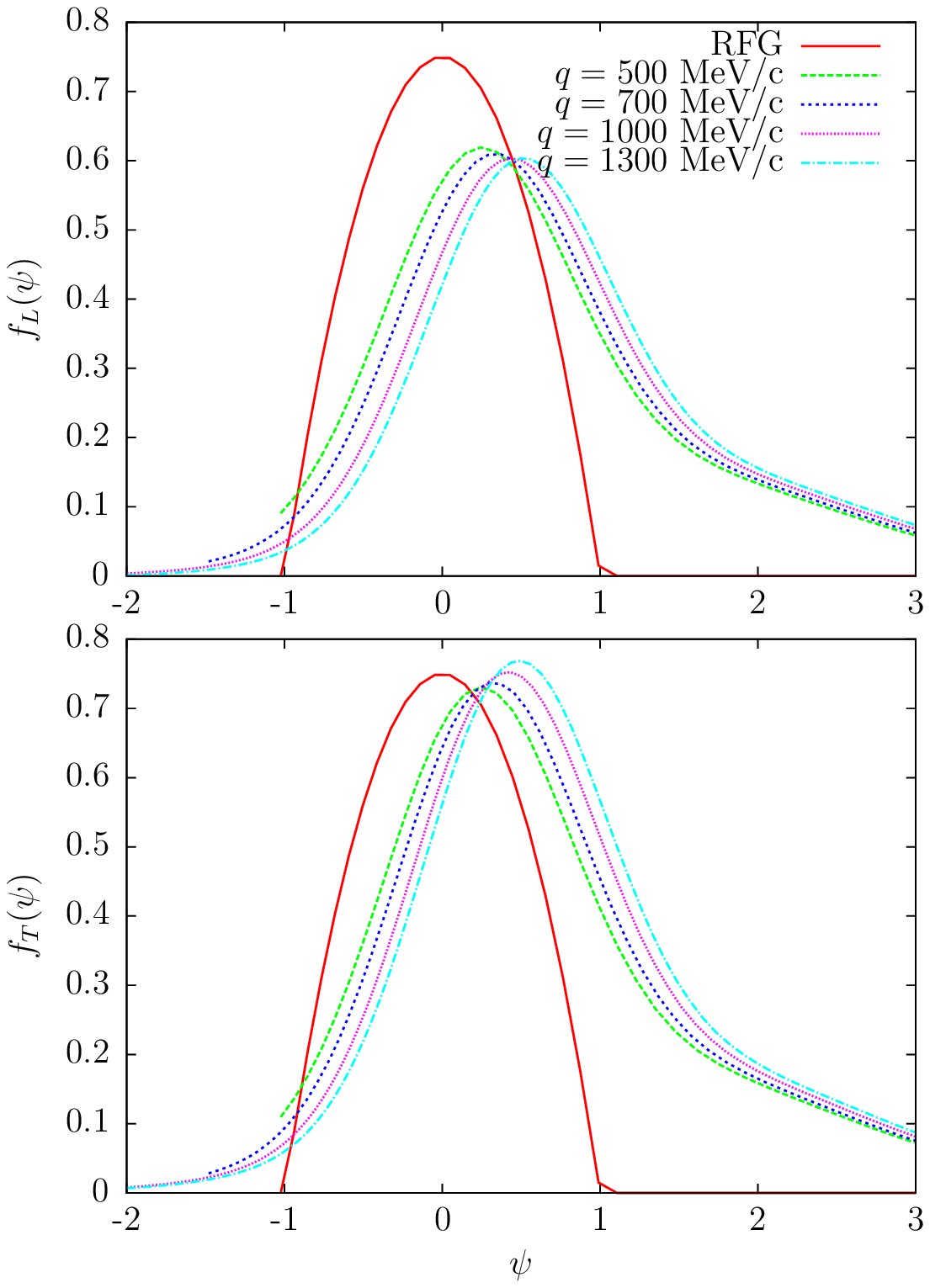}
\caption{ Traditional scaling analysis of the longitudinal and
  transverse scaling functions of $^{12}$C in the SuSAM model, for
  several values of the momentum transfer.  The universal relativistic
  Fermi gas scaling function is also shown for comparison }
\label{scaling}
\end{figure}

\begin{figure}
\includegraphics[width= 8cm, bb=145 390 415 780]{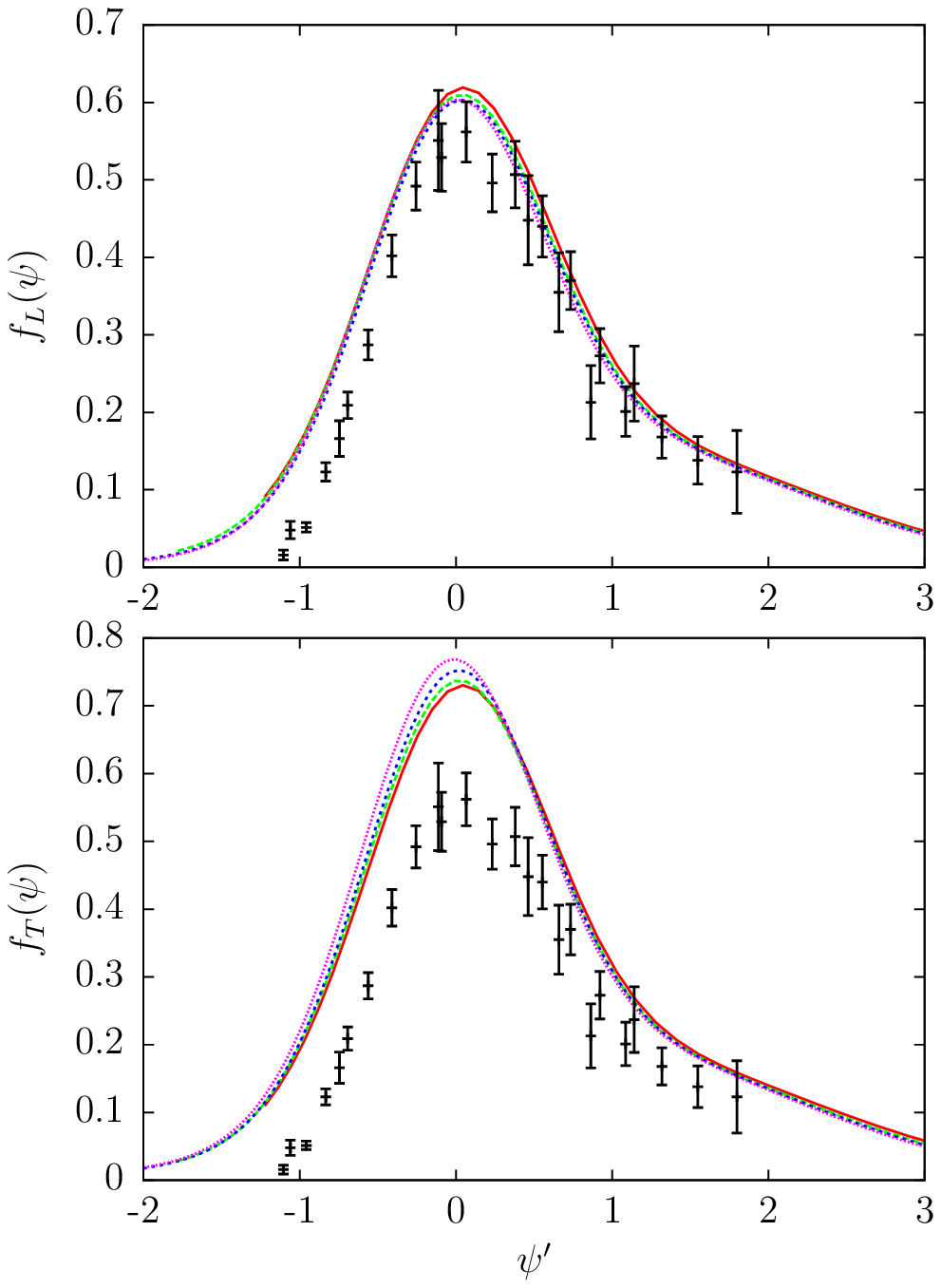}
\caption{ Traditional scaling analysis of the longitudinal and
  transverse scaling functions of $^{12}$C in the SuSAM model, for
  several values of the momentum transfer. They are plotted as a
  function of the shifted variable $\psi'$ so that the maximum in each
  curve is reached at $\psi'=0$. For comparison we also plot the data
  of the longitudinal scaling function obtained from the
  experimental $R_L(q,\omega)$ data in ref. \cite{Mai02}.  }
\label{scaling-exp}
\end{figure}

\begin{figure*}
\includegraphics[width= 14cm, bb=25 270 540 780]{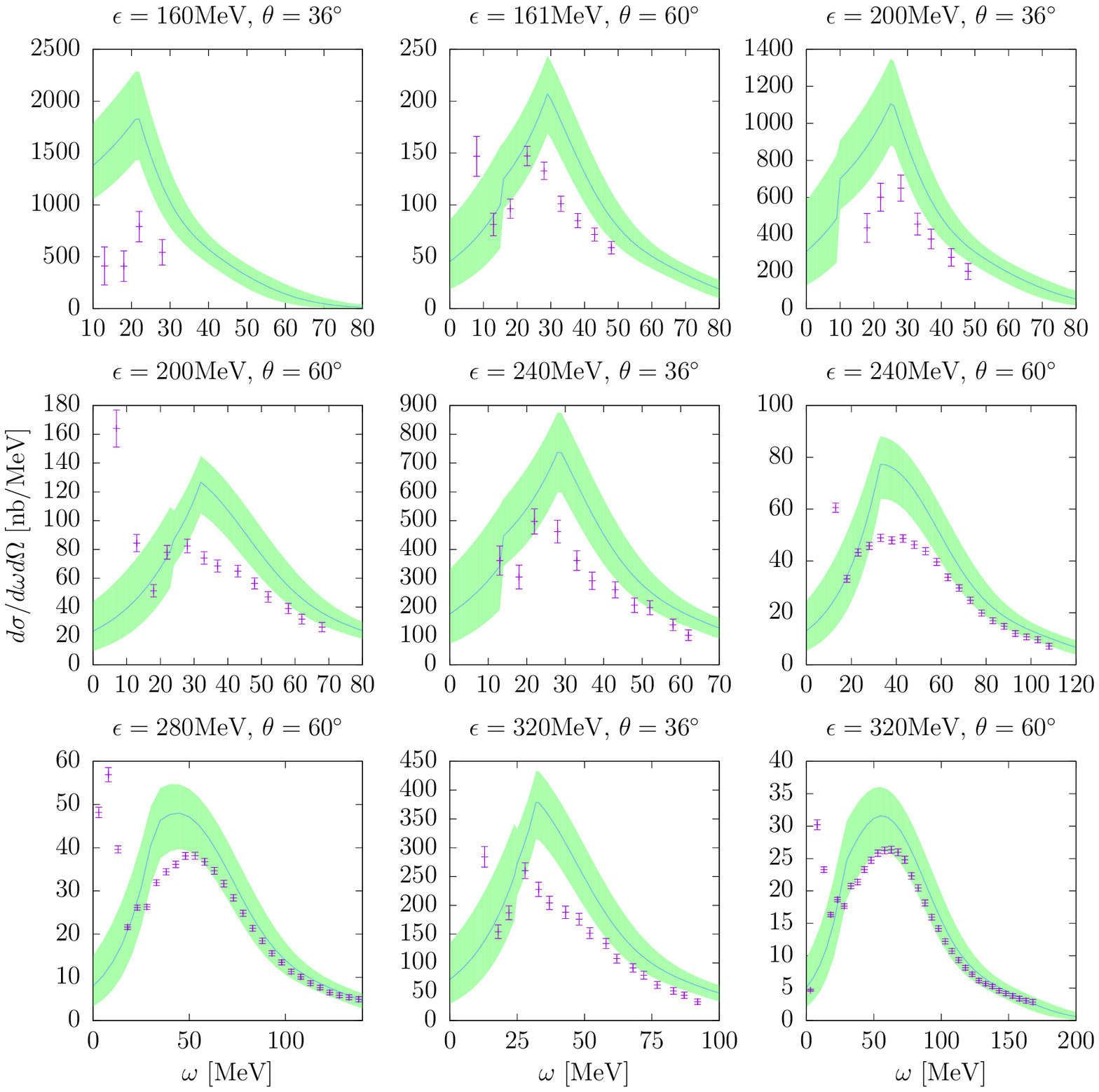}
\caption{ SuSAM predictions and uncertainty 
bands for the quasielastic $(e,e')$ cross section
for several kinematics compared to the experimental data.
 }
\label{valida1}
\end{figure*}

\begin{figure*}
\includegraphics[width= 14cm, bb=25 270 540 780]{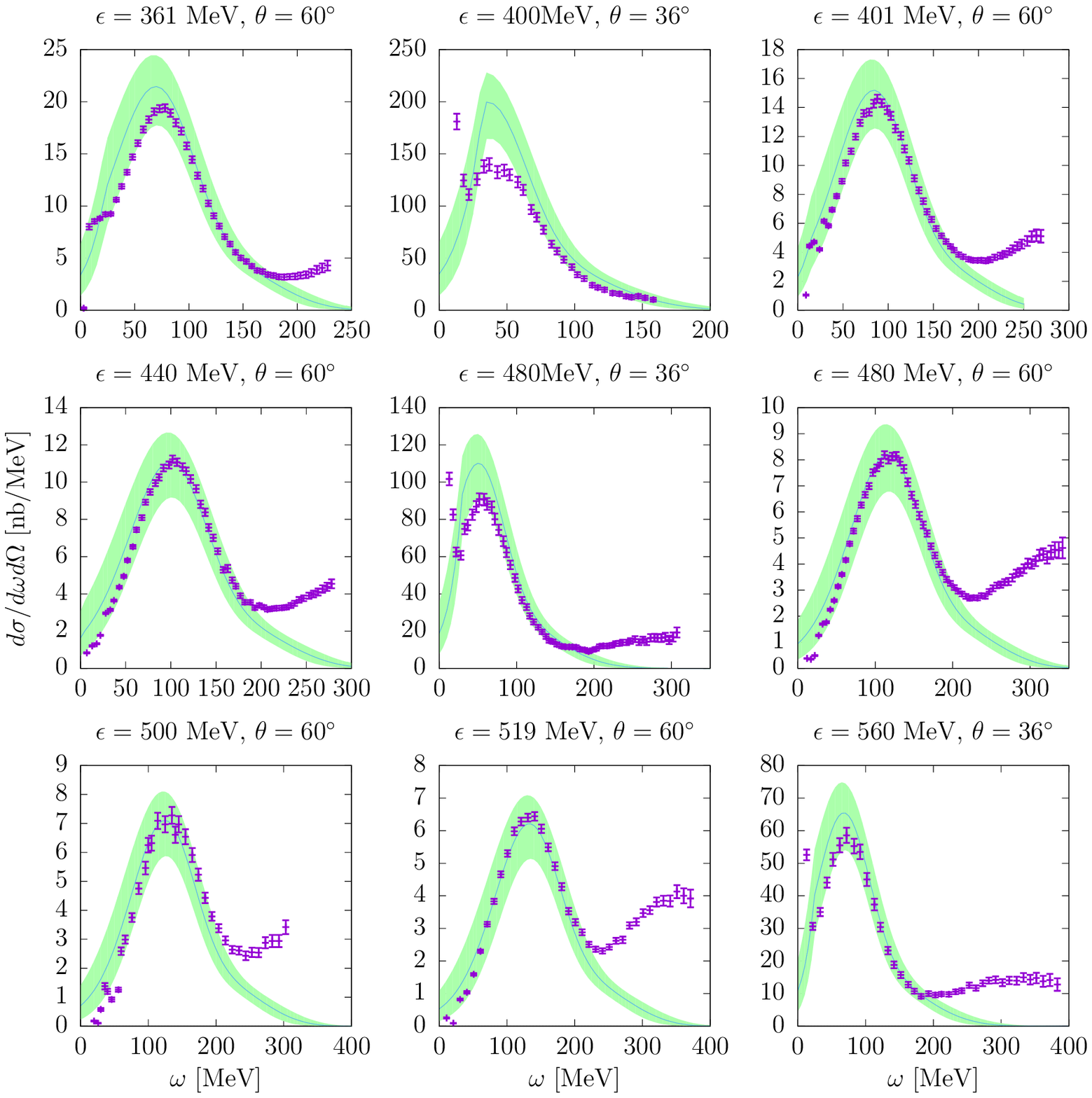}
\caption{ SuSAM predictions and uncertainty 
bands for the quasielastic $(e,e')$ cross section
for several kinematics compared to the experimental data.
 }
\label{valida2}
\end{figure*}

\begin{figure*}
\includegraphics[width= 14cm, bb=25 270 540 780]{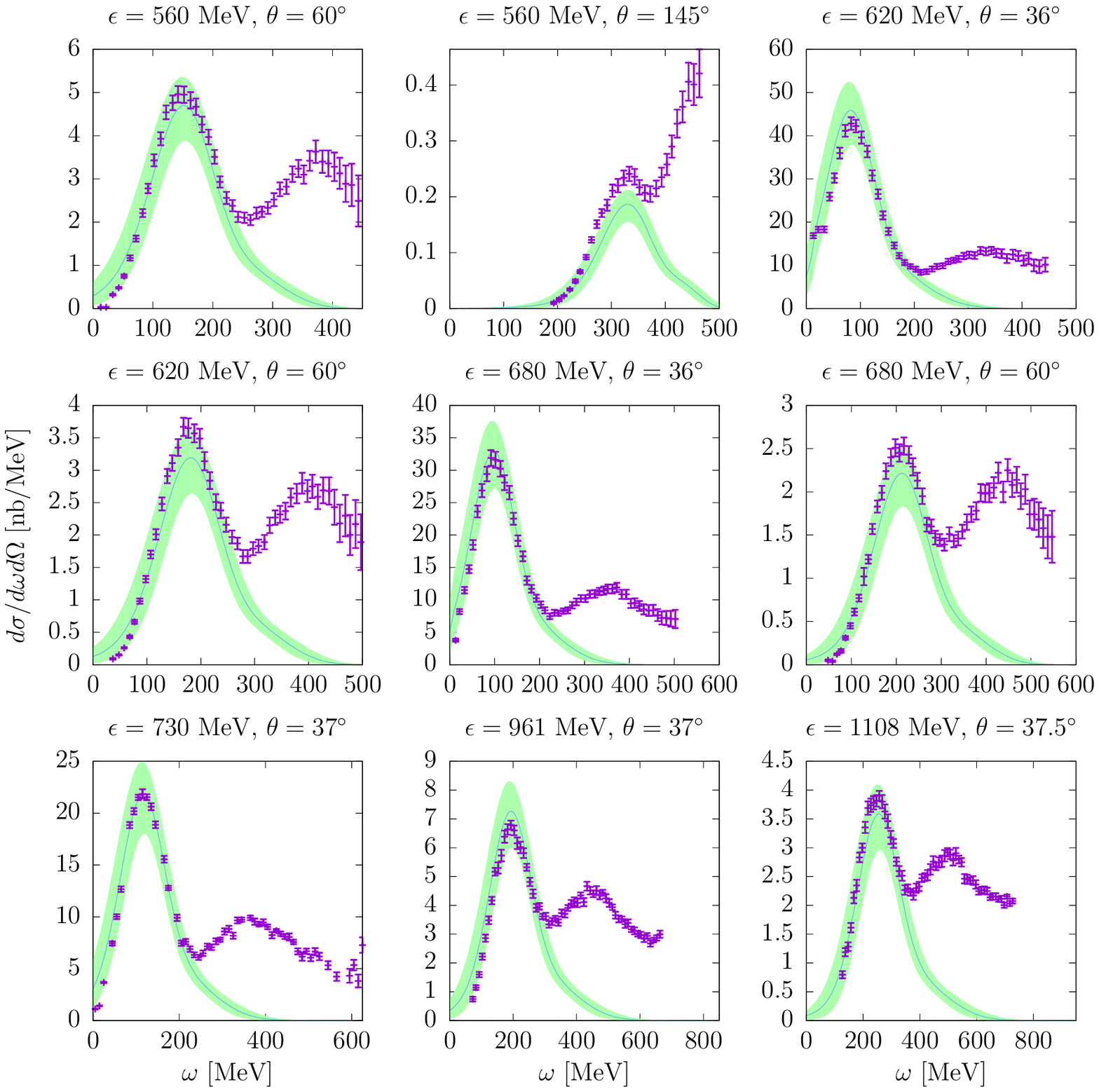}
\caption{ SuSAM predictions and uncertainty 
bands for the quasielastic $(e,e')$ cross section
for several kinematics compared to the experimental data.
 }
\label{valida3}
\end{figure*}

\begin{figure*}
\includegraphics[width= 14cm, bb=25 270 540 780]{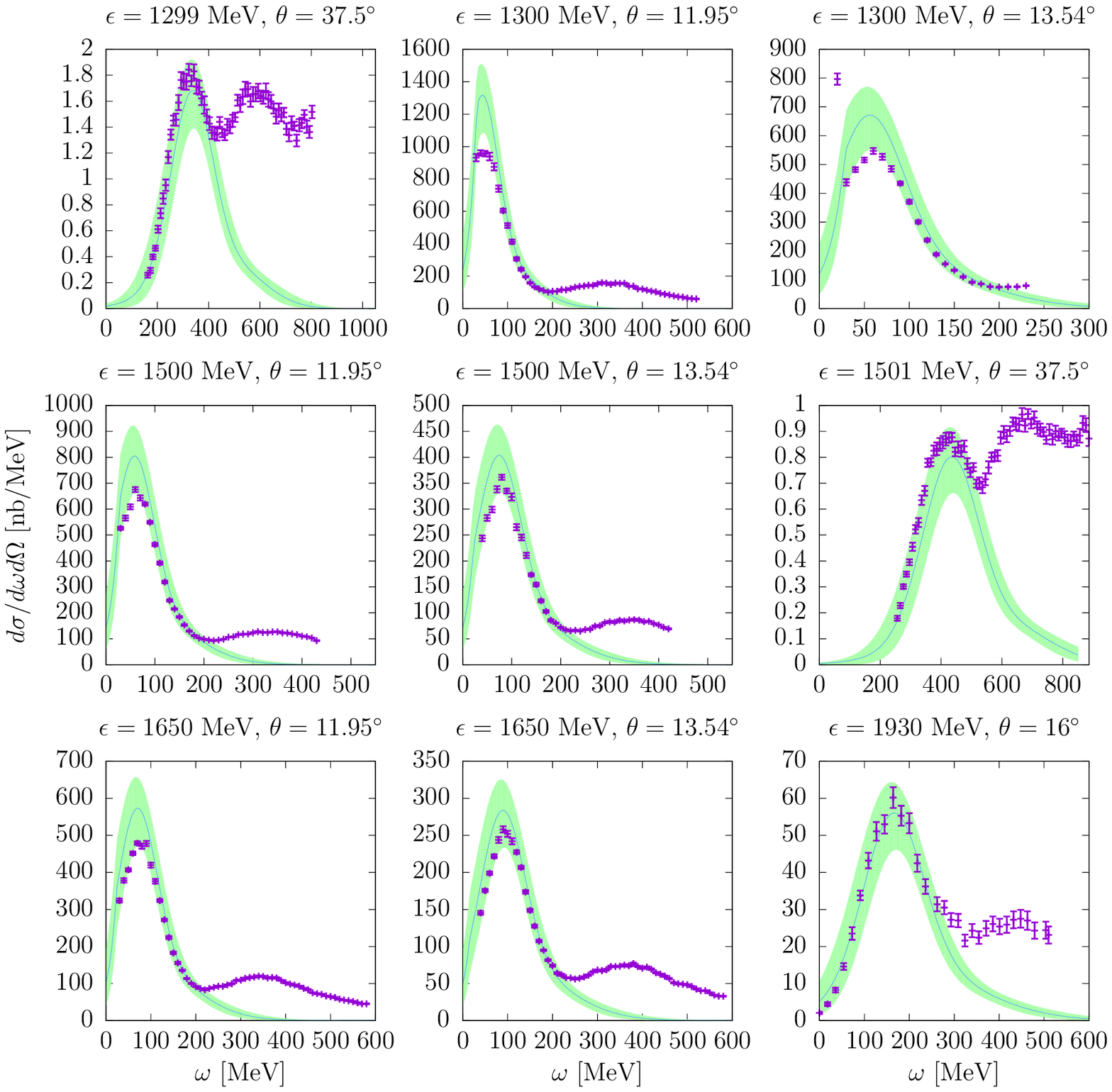}
\caption{ SuSAM predictions and uncertainty 
bands for the quasielastic $(e,e')$ cross section
for several kinematics compared to the experimental data.
 }
\label{valida4}
\end{figure*}

\begin{figure*}
\includegraphics[width= 14cm, bb=25 270 540 780]{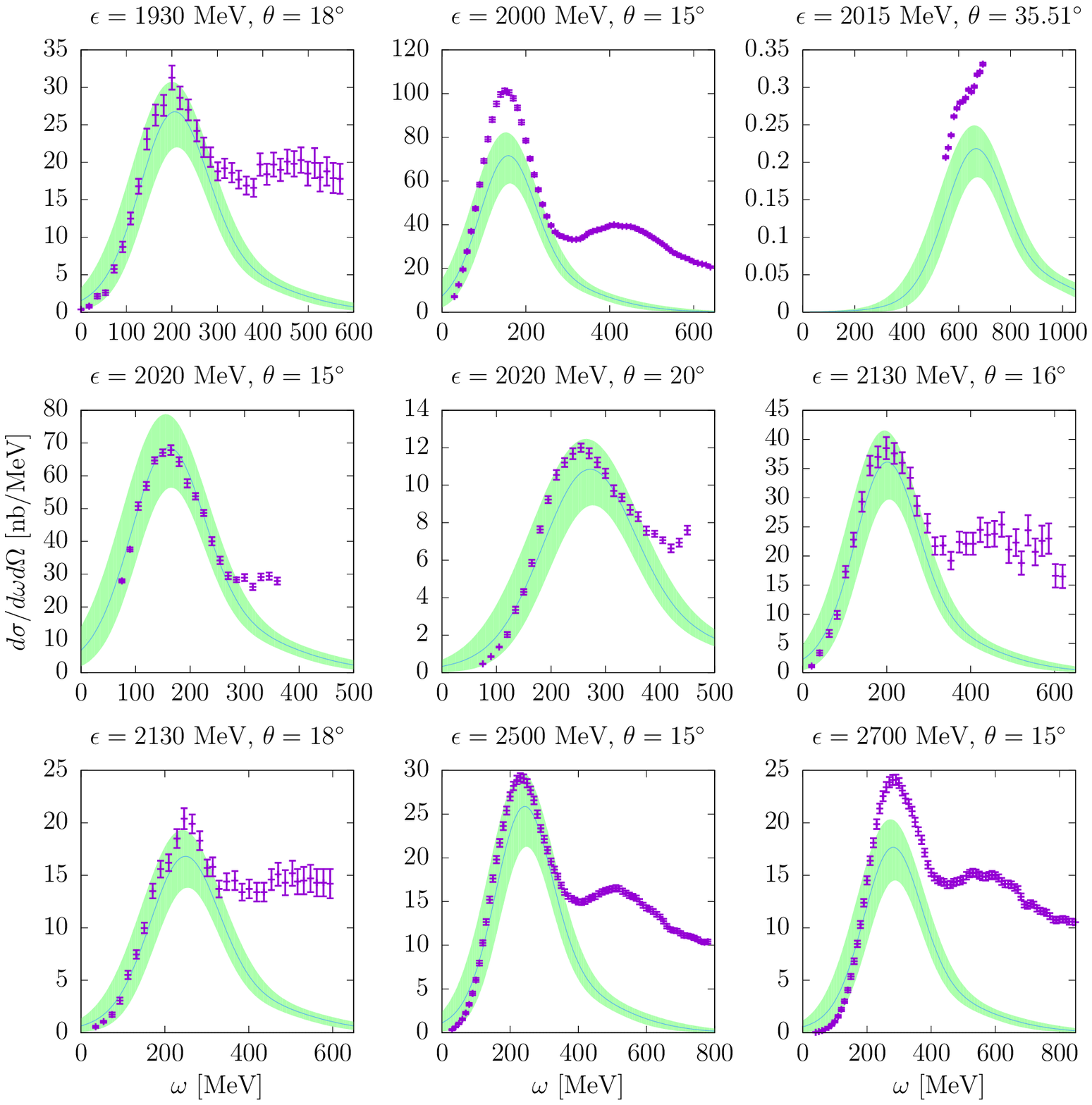}
\caption{ SuSAM predictions and uncertainty 
bands for the quasielastic $(e,e')$ cross section
for several kinematics compared to the experimental data.
 }
\label{valida5}
\end{figure*}

\begin{figure*}
\includegraphics[width= 14cm, bb=25 270 540 780]{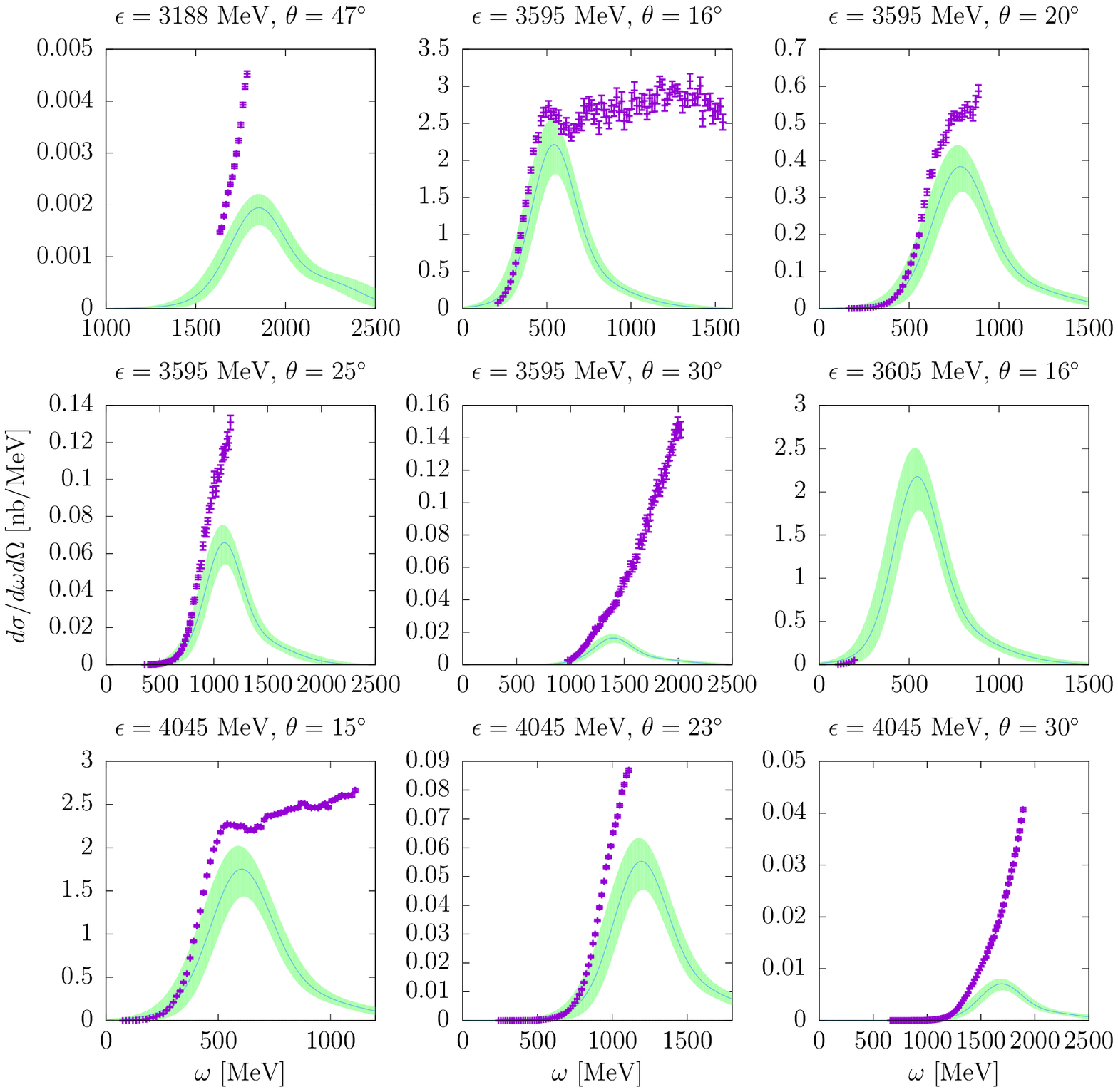}
\caption{ SuSAM predictions and uncertainty 
bands for the quasielastic $(e,e')$ cross section
for several kinematics compared to the experimental data.
 }
\label{valida6}
\end{figure*}

We start with  the
more than 2500 experimental $(e,e')$ cross section data for $^{12}$C
\cite{archive2}.  For every kinematical point we compute the corresponding 
experimental
scaling function $f^*_{\rm exp}$
\begin{equation}
f^*_{\rm exp} =
\frac{\left(\frac{d\sigma}{d\Omega'd\epsilon'}\right)_{\rm exp}}{
  \sigma_{\rm Mott}\left( v_L r_L + v_T r_T \right)}
\end{equation}
In ref. \cite{Ama15} we performed an analysis of the experimental
scaling function for the bulk of data \cite{archive,archive2} by
plotting them against the scaling variable $\psi^*$.  A large fraction
of the data then collapses into a cloud with an asymmetrical 
shape as seen in Fig. 1.  The cloud of data forms a thick band. The
selection of data was made in ref. \cite{Ama15} by measuring the
density of points clustered above a given threshold $n$, inside a
circle of radius $r=0.1$. The selection of data depends on the chosen
value of $n$.  In this work we use $n=25$, meaning that we neglect all
the points with less than 25 neighbours inside a circle of radius
$r=0.1$.  The number of surviving data is around 1000.  The thickness
of the data cloud measures the small degree of scaling violation
around the quasielastic peak. The neglected data correspond to
inelastic excitations and low energy processes that highly violate
scaling and cannot be considered quasielastic processes, as can be
seen in the lower panel of Fig. 1.

\begin{table}
\begin{tabular}{crrrrrr}\hline
   & $a_1$ & $a_2$ & $a_3$ & $b_1$ & $b_2$ & $b_3$ \\ \hline
central 
&-0.0465
& 0.469
& 0.633
& 0.707
& 1.073
& 0.202
\nonumber\\
min
&-0.0270
& 0.442
& 0.598
& 0.967
& 0.705
& 0.149
\nonumber\\
max
& -0.0779
& 0.561
& 0.760
& 0.965
& 1.279
& 0.200
\\ \hline
\end{tabular}
\caption{Parameters of our fit of the phenomenological 
scaling function central value, $f^*(\psi^*)$, 
and of the lower and upper boundaries (min and max, respectively).} 
\end{table}

In Fig. 1 we show also our new parametrization of the scaling function
after a fit to the selected (quasielastic) experimental data. They are
well described as a sum of two Gaussian functions
\begin{equation}\label{fitpar}
f^*(\psi^*)  = 
a_3 e^{-(\psi^*-a_1)^2/(2 a_2^2)}+b_3 e^{-(\psi^*-b_1)^2/(2b_2^2)}
\end{equation}
The coefficients are given in table 1.

The lower and upper limits of the experimental data band have also
been parametrized as sum of two Gaussians, with coefficients
$a_i^{min/max}$, $b_i^{min/max}$, given in table 1 as well.

Our phenomenological scaling function describes the center values of
the selected data cloud and the average thickness. The thickness can
be interpreted as a fluctuation produced by nuclear effects beyond the
impulse approximation (finite size effects, short-range
NN-correlations, long-range RPA, meson-exchange currents, virtual
$\Delta$ excitation, two-particle emission, final state
interaction) \cite{Ama15}.

\section{Results}

In this section we use the new phenomenological scaling function of
the SuSAM* model, given by eq. (\ref{fitpar}), to compute the response
functions and cross section of $^{12}$C and the corresponding
uncertainty band derived from the thickness of the fitted data set.

In figs 2 and 3 we show the longitudinal and transverse response
functions for several values of the momentum transfer. They are
compared to the free RFG and to the interacting RFG with
$M^*=0.8$. The effect of the effective mass is a shift of the
responses to higher energies, because the position of the quasielastic
peak is given by $\omega= \sqrt{q^2+m_N^*{}^2}-m_N^*$.  Note that this
shift gives the correct position of the quasielastic peak without need
of introducing a separation energy parameter \cite{Ros80,Weh93}. 
In fact the position of the 
peak with the SuSAM* model almost coincides with the RFG with $M^*=0.8$.

The introduction of the phenomenological scaling function produces a
prominent tail for high energy transfer, which extends much high that
the upper end of the RFG responses, which is more in accordance with
the experimental data. This high end is a consequence of the maximum
momentum for the nucleons in the Fermi gas, bound by the Fermi
momentum, corresponding to $\psi^*=1$. Since the data in Fig.1 extends
above $\psi^*=1$, this indicates the presence of final state
interactions and also high momentum components in the nuclear wave
function in the ground state, which are thought to be caused by finite
size effects and nuclear correlations \cite{Wir14,Rui16}. Another
smaller tail appears for low energy also produced by nuclear effects.

The structure of the high energy tail is different in the longitudinal
and transverse responses. The longitudinal response presents a
prominent shoulder, which is not so prominent in the transverse one,
although both responses have been calculated with the same
phenomenological scaling function. Given that the scaling function
does not present a similar shoulder, one can trace back its origin to
the energy dependence of the single nucleon responses in the medium,
that have been calculated with $M^*=0.8$, see
Eqs. (\ref{single}--\ref{GM}).

In comparing the ratio between $R_T$ and $R_L$ from figs. 2 and 3 one
can also observe an enhancement for the case $M^*=0.8$. This is
related with the known
enhancement of the lower components in the relativistic
spinor in the nuclear medium.

In Figs 4 and 5 we show the longitudinal and transverse scaling
functions $f_L(\psi)$ and $f_T(\psi)$, obtained form the response of
Fig.s 2, 3, by dividing by the single nucleon responses with effective
mass $M^*=1$. These can be directly compared to the traditional SuSA
scaling functions. They are plotted against the original scaling
variable $\psi$ for different values of $q$. It is apparent that they
are almost independent on $q$. They are compared to the scaling function of
the free RFG. The $f_L(\psi)$ scaling function is smaller than the
free one as in the case of the phenomenological SuSA scaling function.
The $f_T(\psi)$ scaling function is larger and almost of the same size
as the free one, being this again a consequence of 
the enhancement of the transverse response function.

The $\psi$-scaling properties of the SuSAM* model can be better 
observed in
fig. 6.  The scaling is only approximate as it happens with the RMF
model. There is a shift with respect to the RFG parabola, which
increases slightly with the momentum transfer. This increase is approximately
linear.  The enhancement of the transverse response also increases with
$q$ (bottom panel of Fig. 6).

In fig. 7 we compare the $\psi$-scaling function of our model with the
experimental data of the longitudinal scaling function $f_L(\psi)$
obtained in ref. \cite{Mai02}.  The longitudinal scaling function is
well reproduced, except a slight disagreement for low $\psi'$, because we
fit the cross section and not $R_L$. The transverse scaling function
in our model is evidently larger than the experimental $f_L(\psi')$
in an amount about 20 \%, similar to the RMF results of \cite{Cab07}.

In figs. 8--13 we show the predictions of our model for the $(e,e')$
cross section compared to the experimental data.  Our global
description is quite acceptable given the few parameters of the SuSAM*
model.  A large fraction of the data fall inside our uncertainty
band. In fact, most of the data used to perform the fit, and
displayed in Fig. 1 (top) are inside our prediction bands by
construction.  The data that lie outside our prediction bands are
those clearly in the inelastic or deep region and those corresponding
to low excitation energy, and therefore break $\psi^*$-scaling because
they fall outside the quasielastic region defined in fig. 1 (top).

\section{Conclusions}

In this paper we have investigated a novel scaling approach based on a
new scaling variable $\psi^*$ extracted from the scaling properties of
the RMF model in nuclear matter. Within this model we have obtained a
phenomenological scaling function $f^*(\psi^*)$ from the inclusive
(e,e') reaction data off $^{12}$C, after a selection procedure of the
quasielastic subset of data.  This new phenomenological scaling
function has been parametrized as the sum of two Gaussians.
Additionally, an uncertainty band has been assigned to the
phenomenological scaling function from the dispersion of the data set
around the central value.

With this scaling function we have calculated the longitudinal and
transverse response functions and the conventional scaling functions
of the SuSA model.  Our model contains the enhancement of the
transverse components of the electromagnetic current by construction. 
This confirms
the RMF interpretation of the enhancement of the transverse response
in terms of the relativistic modification of the lower
components of the nucleon spinors in the medium, that we encode with
the effective nucleon mass reduction of $M^*=0.8$.

Finally we have computed the differential quasielastic cross section
with an uncertainty band generated by the scaling function thickness
and compared to the world $^{12}$C$(e,e')$ data, with a reasonable
description of around one thousand data, which thus 
can be tagged as truly quasielastic.

This model predicts a quasielastic cross section directly from the
data without any theoretical assumption, besides the requirements of
gauge invariance, relativity and scaling, which determines the values
of the relativistic effective mass and the Fermi momentum. The rest of
nuclear effects contributing to the quasielastic reaction, have been
encoded into the parametrized scaling function $f^*(\psi^*)$ within an
uncertainty band.  Any model aiming to describe the quasielastic cross
section at intermediate energies should lie inside the SuSAM*
uncertainty band. Therefore, our scaling function parametrization
provides a novel test for theoretical scaling studies. This imposes
constraints over the transverse enhancement, additional to those imposed by 
the longitudinal scaling function in the SuSA model.
In line with the current revival of electron scattering this model can
be easily extended to provide tight constraints in quasielastic
neutrino scattering.

\section{Acknowledgements}

This work is supported by Spanish DGI (grant FIS2014-59386-P) and
Junta de Andalucia (grant FQM225). I.R.S. acknowledges support from
the Ministerio de Economia y Competitividad 
(grant Juan de la Cierva-Incorporacion).



\begin{thebibliography}{99}


\bibitem{Nomad09} V Lyubushkin et al.   (NOMAD Collaboration),
 Eur. Phys. J. C 63 (2009), 355.


\bibitem{Agu10} A. Aguilar-Arevalo {\em et al.} (MiniBooNE Collaboration),
Phys. Rev. D {\bf 81}, 092005 (2010).

\bibitem{Agu13} A. Aguilar-Arevalo {\em et al.} (MiniBooNE Collaboration),
Phys. Rev. D {\bf 88}, 032001 (2013).

\bibitem{Fio13} G.A. Fiorentini {\em et al.} (MINERvA Collaboration)
Phys. Rev. Lett. {\bf 111}, 022502 (2013).


\bibitem{Abe13} K. Abe {\em et al.}, (T2K Collaboration),
Phys. Rev. D {\bf 87}, 092003 (2013).

\bibitem{Lov16}
A. Lovato, S. Gandolfi, J. Carlson, Steven C. Pieper, R. Schiavilla, 
 Phys.Rev.Lett. 117 (2016) 082501.


\bibitem{Ank15} A. M. Ankowski, O. Benhar, M. Sakuda, Phys. Rev. D 91,
  033005 (2015).


\bibitem{Roc16}
N. Rocco, A. Lovato, O. Benhar, 
    Phys.Rev.Lett. 116 (2016)  192501.



\bibitem{Pan15} 
  V.~Pandey, N.~Jachowicz, T.~Van Cuyck, J.~Ryckebusch and M.~Martini,
  Phys.\ Rev.\ C {\bf 92}, no. 2, 024606 (2015).

\bibitem{Ama02} 
  J.~E.~Amaro, M.~B.~Barbaro, J.~A.~Caballero, T.~W.~Donnelly and A.~Molinari,
  Phys.\ Rept.\  {\bf 368}, 317 (2002).

\bibitem{Ama05} 
  J.~E.~Amaro, M.~B.~Barbaro, J.~A.~Caballero, T.~W.~Donnelly and C.~Maieron,
  Phys.\ Rev.\ C {\bf 71}, 065501 (2005).

\bibitem{Cab07} 
  J.~A.~Caballero, J.~E.~Amaro, M.~B.~Barbaro, T.~W.~Donnelly and J.~M.~Udias,
  Phys.\ Lett.\ B {\bf 653}, 366 (2007).


\bibitem{Udi99} 
  J.~M.~Udias, J.~A.~Caballero, E.~Moya de Guerra, J.~E.~Amaro and T.~W.~Donnelly,
  Phys.\ Rev.\ Lett.\  {\bf 83}, 5451 (1999).


\bibitem{Bod14} A. Bodek, M.E. Christy, and B. Coopersmith, 
Eur. Phys. Jou. C 74, 3091 (2014). 

\bibitem{Gil97}
A.~Gil, J.~Nieves and E.~Oset,
Nucl.\ Phys.\ A  {\bf 627} (1997) 543.


\bibitem{Sim16} 
  I.~Ruiz Simo, J.~E.~Amaro, M.~B.~Barbaro, A.~De Pace, J.~A.~Caballero and T.~W.~Donnelly,
  arXiv:1604.08423 [nucl-th].


\bibitem{Nie17} J. Nieves, J.E. Sobczyk,  arXiv:1701.03628 [nucl-th].


\bibitem{Don99a} 
  T.~W.~Donnelly and I.~Sick,
  Phys.\ Rev.\ Lett.\  {\bf 82}, 3212 (1999).

\bibitem{Don99b} 
  T.~W.~Donnelly and I.~Sick,
  Phys.\ Rev.\ C {\bf 60}, 065502 (1999).



\bibitem{Alb88} W.M. Alberico, A. Molinari, T.W. Donnelly, E. L. Kronenberg, 
and J.W. Van Orden, Phys Rev. C 38 (1988) 1801.


\bibitem{Mai02} 
  C.~Maieron, T.~W.~Donnelly and I.~Sick,
  Phys.\ Rev.\ C {\bf 65}, 025502 (2002).

\bibitem{Ama04} 
  J.~E.~Amaro, M.~B.~Barbaro, J.~A.~Caballero, T.~W.~Donnelly, A. Molinari, I. Sick,
Phys. Rev. C {\bf 71}, 015501 (2005).


\bibitem{Hor91} C.J. Horowitz, D.P. Murdock, and B.D. Serot,
in {\em Computational Nuclear Physics Vol. 1}, Springer-Verlag, Berlin 1991.


\bibitem{For83} 
  T.~De Forest,
  Nucl.\ Phys.\ A {\bf 392}, 232 (1983).


\bibitem{Gon14} R. Gonzalez-Jimenez, G.D. Megias, M.B. Barbaro, J.A. Caballero, 
and T.W. Donnelly, Phys. Rev. C 90, 035501 (2014).

\bibitem{Ama15} 
  J.~E.~Amaro, E.~Ruiz Arriola and I.~Ruiz Simo,
  Phys.\ Rev.\ C {\bf 92}, no. 5, 054607 (2015)

\bibitem{Ros80} R. Rosenfelder, Ann. Phys. (N.Y:) 128, 188 (1980)

\bibitem{Ser86} B.D. Serot, and J.D. Walecka, Adv. Nucl. Phys. 16 (1986) 1.

\bibitem{Bar98} M.B. Barbaro, R. Cenni, A. De Pace, T.W. Donnelly,
  A. Molinari, Nucl. Phys. A 643 (1998) 137.

\bibitem{archive2} O. Benhar, D. Day, and I. Sick, 
http://faculty.virginia.edu/qes-archive/  

\bibitem{archive} O. Benhar, D. Day and I. Sick, arXiv:nucl-ex/0603032.

\bibitem{Ben08} O. Benhar, D. Day, and I. Sick, Rev Mod Phys. 80 (2008) 189

\bibitem{Weh93} K. Wehrberger, Phys. Rep. 225 (1993) 273.

\bibitem{Wir14} 
  R.~B.~Wiringa, R.~Schiavilla, S.~C.~Pieper and J.~Carlson,
  Phys.\ Rev.\ C {\bf 89}, no. 2, 024305 (2014)

\bibitem{Rui16} 
  I.~Ruiz Simo, R.~N.~Perez, J.~E.~Amaro and E.~Ruiz Arriola,
  arXiv:1612.06228 [nucl-th].


\end{thebibliography}
\end{document}